%% file: neurips_2025.tex
\documentclass{article}

\usepackage[sglblindworkshop, final]{ai4nextg_neurips_2025}

\usepackage[utf8]{inputenc} 
\usepackage[T1]{fontenc}    
\usepackage{hyperref}       
\usepackage{url}            
\usepackage{booktabs}       
\usepackage{amsfonts}       
\usepackage{nicefrac}       
\usepackage{microtype}      
\usepackage{xcolor}         
\usepackage{natbib}
\usepackage{balance} 
\usepackage{xcolor}
\usepackage{algorithm}
\usepackage{algpseudocode}
\usepackage{tcolorbox} 
\usepackage{amsmath} 
\usepackage{algorithm}
\usepackage{algorithmicx}  
\usepackage{algpseudocode}
\usepackage{setspace}
\usepackage{enumitem}
\usepackage{hyperref}
\usepackage{wrapfig}  

\usepackage{graphicx}
\usepackage{booktabs}
\usepackage{multirow}
\usepackage{subcaption} 
\usepackage{multirow} 
\usepackage{graphicx,booktabs,multirow,subcaption}
\usepackage{caption}
\newcommand{\BibTeX}{\rm B\kern-.05em{\sc i\kern-.025em b}\kern-.08em\TeX}

\definecolor{delayed}{HTML}{dfe8e0} 
\definecolor{llm}{HTML}{a8c7f0} 
\definecolor{marl}{HTML}{f4b78a} 
\definecolor{dreward}{HTML}{8EAD91}
\definecolor{feedback}{HTML}{e2c0da}
\definecolor{align}{HTML}{dad5a4}
\newcommand{\framework}{{AURA }}

\definecolor{mycitecolor}{HTML}{c664a0}

\hypersetup{
  colorlinks=true,     
  linkcolor=black,     
  citecolor=mycitecolor, 
  urlcolor=black       
}

\title{AURA: Adaptive Unified Reasoning and Automation 
with LLM-Guided MARL for NextG Cellular Networks}

\author{%
  Narjes Nourzad \\ 
  Department of Electrical and Computer Engineering\\
   University of Southern California\\
  Los Angeles, CA 90007 \\
  \texttt{nourzad@usc.edu} \\
  \And
  Mingyu Zong \\
  Department of Computer Science\\
  University of Southern California \\
  Los Angeles, CA 90007 \\
  \texttt{mzong@usc.edu} \\
  \AND
  Bhaskar Krishnamachari \\
  Department of Electrical and Computer Engineering \\ 
  Department of Computer Science\\
  University of Southern California \\
  Los Angeles, CA 90007 \\
  \texttt{bkrishna@usc.edu} \\
}

\begin{document}

\maketitle

\begin{abstract}
Next-generation (NextG) cellular networks are expected to manage dynamic traffic while sustaining high performance. Large language models (LLMs) provide strategic reasoning for 6G planning, but their computational cost and latency limit real-time use. Multi-agent reinforcement learning (MARL) supports localized adaptation, yet coordination at scale remains challenging.
We present \textsc{AURA}, a framework that integrates cloud-based LLMs for high-level planning with base stations modeled as MARL agents for local decision-making. The LLM generates objectives and subgoals from its understanding of the environment and reasoning capabilities, while agents at base stations execute these objectives autonomously, guided by a trust mechanism that balances local learning with external input. To reduce latency, \framework employs batched communication so that agents update the LLM’s view of the environment and receive improved feedback.
In a simulated 6G scenario, \framework improves resilience, reducing dropped handoff requests by more than half under normal and high traffic and lowering system failures. Agents use LLM input in fewer than 60\% of cases, showing that guidance augments rather than replaces local adaptability, thereby mitigating latency and hallucination risks.
These results highlight the promise of combining LLM reasoning with MARL adaptability for scalable, real-time NextG network management.

\end{abstract}

\input{introduction}

\input{methodology}

\input{eval}

\input{exp_result}

\small
\bibliography{references}
\bibliographystyle{plainnat}

\input{appendix}
\input{related_work}

\input{alg}
\input{exp_setup}

\end{document}

%% file: introduction.tex
\section{Introduction}

6G cellular networks are expected to provide high data transmission speeds and seamless connectivity, supporting devices from autonomous vehicles to personal gadgets at large-scale events~\citep{banafaa20236g, chataut20246g}. 
These advancements have the potential to redefine how we connect and communicate, but this promise has yet to translate into practical, deployable solutions~\citep{maduranga2024ai, shahjalal2023enabling, cui2024overview}. 
Next-generation networks face increasing challenges in managing dynamic and unpredictable traffic.
High user density, coupled with ultra-low latency demands for applications such as holographic imaging and haptic communication, requires sophisticated algorithms~\citep{maduranga2024ai, dogra2023intelligent}.   Consequently, the imperative to balance reliability, energy efficiency, high data rates, and low latency has led researchers to envision 6G as AI-driven networks. AI integration aims to improve throughput, reliability, and latency, while enabling self-optimizing operations through intelligent resource allocation and adaptive traffic management~\citep{ yang2020artificial, noman2023machine}.

Traditional heuristic methods fall short in meeting these demands due to their limited scalability and difficulty in quantifying performance gaps~\citep{abasi2024metaheuristic}, prompting a shift toward machine learning (ML) methods~\citep{kim2023towards, wang2023artificial}.
ML and deep learning (DL) algorithms have considerably improved network performance, but they face major challenges in real-world 6G settings, because of their reliance on annotated data. The NP-hard nature of labeling makes supervised learning impractical for dynamic settings, lacking both scalability and adaptability~\citep{cui2024overview}. Reinforcement learning (RL) offers an alternative to supervised methods by eliminating the need for labeled data.
Multi-agent reinforcement learning (MARL) extends RL to environments with multiple agents that must coordinate or compete~\citep{bhati2023curriculum}. By enabling localized decision-making, MARL provides scalability and robustness~\citep{sun2023hmaac}, making it a natural fit for 6G network management~\citep{chu2019multi}.
However, current MARL algorithms struggle to learn distributed policies for cooperative tasks, particularly in sparse-reward, dynamic environments with large action spaces, characteristics of next-generation networks~\citep{feriani2021single}.
\begin{wrapfigure}{r}{0.6\textwidth}   
\includegraphics[width=0.6\textwidth]{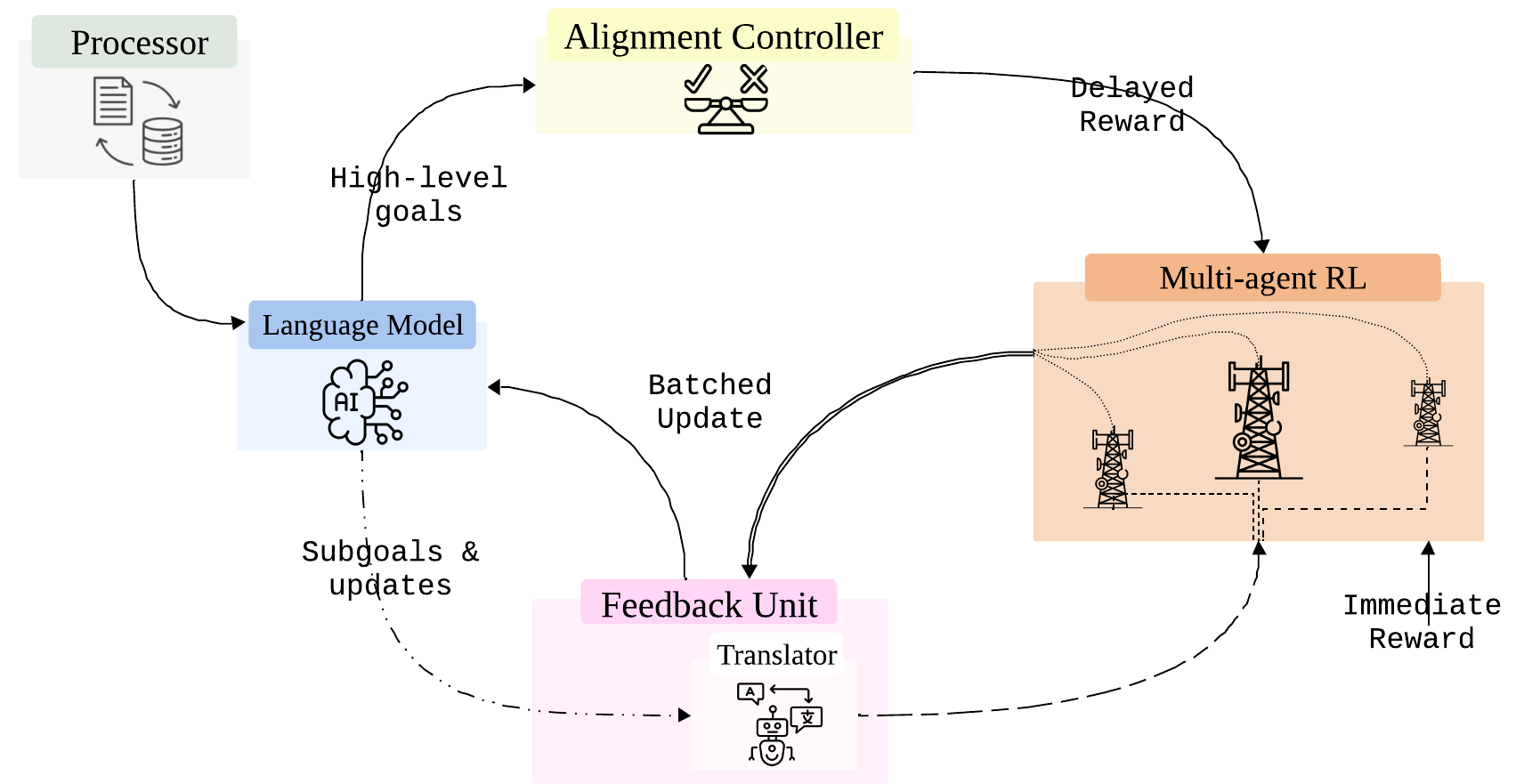}
    \caption{Illustration of the \textit{AURA} Architecture. 
    A \textcolor{llm}{cloud-based LLM} sets high-level objectives from unified inputs processed by the multimodal encoder. 
    Distributed \textcolor{marl}{MARL agents} adapt locally through actionable subgoals, 
    while the \textcolor{align}{aliment controller} aligns policies and assigns rewards. 
    Iterative \textcolor{feedback}{feedback} refines decisions for dynamic network management.}
    \label{fig:aura-model}
\end{wrapfigure}

 \vspace{-12pt}
Centralized training with decentralized execution (CTDE) is widely adopted to overcome the limitations of independent learning~\citep{wang2023dm2}. Nevertheless, CTDE itself faces constraints such as limited agent communication, difficulties adapting to non-stationary environments, and scalability issues as the number of agents grows~\citep{oroojlooy2023review}.  
Additionally, many MARL-based solutions suffer from high complexity, as agents must encode large amounts of information into their policies. Because these policies are rarely generalizable, they are often trained from scratch, further increasing computational costs and slowing convergence~\citep{yang2025llm}.
These limitations suggest the need for alternative approaches that combine high-level reasoning with localized adaptability to ensure cohesive multi-agent coordination.
In parallel, large language models (LLMs) with billions of pre-trained parameters have demonstrated remarkable capabilities in reasoning, planning, and structured decision-making~\citep{xi2025rise}.  Their ability to generalize across tasks makes them a compelling tool for network intelligence. In particular, LLMs can function as high-level semantic planners, leveraging in-context learning and prior knowledge~\citep{ahn2022pretraining}. 
Through planning, LLMs decompose high-level goals into actionable low-level tasks; through reasoning, they structure complex problems into priors and beliefs. However, directly deploying LLMs for fine-grained, real-time network parameter adaptation remains infeasible due to their computational overhead and latency.
To reconcile these complementary strengths and weaknesses, we propose \textit{AURA: Adaptive Unified Reasoning and Automation} for NextG cellular networks.

\textit{AURA} is a hierarchical framework that combines the predictive capabilities of Large Language Models (LLMs) with the real-time adaptability of multi-agent reinforcement learning (MARL).
It addresses core 6G challenges such as dynamic traffic patterns, fluctuating user demands, and evolving network conditions. 
A multimodal processor unifies diverse data sources, which the LLM uses to set high-level objectives and subgoals through techniques like Chain-of-Thought reasoning~\citep{wei2022chain} and pre-trained policies. 
Each base station acts as a MARL agent that autonomously executes its objectives under a trust mechanism, balancing local learning with strategic oversight. A centralized controller coordinates these local actions with global goals through structured rewards.
Lightweight communication and batched feedback further enhance scalability and responsiveness, drawing inspiration from reinforcement learning from AI feedback (RLAIF)~\citep{bai2022constitutional, leerlaif}.  
We evaluate \framework in a 6G networks operating in a scenario with dynamic user demands. 
The LLM anticipates congestion and generates preemptive objectives, while MARL agents adapt in real time, taking actions adjusting transmit power and enabling handoffs. 
Preliminary results show that \textsc{AURA} significantly reduce dropped requests and system failures compared to pure MARL baselines, with the largest gains under normal and high traffic. Notably, \textsc{AURA} achieves these improvements while relying on LLM input in fewer than 60\% of cases. This is desirable because frequent LLM queries introduce latency, create a central point of failure, and risk propagating hallucinated or suboptimal suggestions. Since LLMs cannot directly interact with the environment or receive real-time feedback, overreliance would limit adaptability.

%% file: methodology.tex
\section{Methodology}
We now introduce \textit{AURA}, an Adaptive Unified Reasoning and Automation framework. Our design is guided by two key objectives: leveraging LLMs for high-level reasoning and ensuring computationally efficient, real-time decision-making via MARL.
The following subsections outline \textit{AURA}’s model structure  (see Figure~\ref{fig:aura-model}) and its role in AI-driven network management.

 \noindent {\textbf{Multimodal Processor for Encoding.}}
The multimodal processor at the LLM input integrates diverse data from modern networks, including time-series metrics, statistical logs, and external sources such as social media, GIS data, and historical traffic. 
GIS inputs, for example, capture spatial patterns like node distribution or congestion. 
Since standard LMs are limited to text, the processor transforms all heterogeneous inputs into a shared latent representation aligned with language tokens, preserving numerical, spatial, and textual information for LLM analysis.

 \noindent {\textbf{LLM for High-Level Planning.}}  
The centralized LLM integrates processed representations to predict demand surges and guide both immediate resource allocation and long-term planning. 
It follows a two-tiered strategy: for common scenarios (e.g., typical congestion), it selects from a repository of pre-trained policies (“offline playbook”); for novel or complex conditions, it applies semantic reasoning (e.g., Chain-of-Thought~\citep{wei2022chain}) to adjust reward structures or environment parameters for MARL adaptation.  
Rather than micromanaging interactions, the LLM defines dynamic subgoals and rewards, enabling MARL agents to implement responses locally. 
This delegation ensures that global objectives are realized through policies that remain flexible and adaptive at the local level~\citep{zhuang2024yolo}.

 \noindent {\textbf{MARL Agents for Local Adaptation.}}  
At individual base stations, MARL agents adjust parameters (e.g., power) using real-time environmental feedback. 
They refine policies through trial-and-error and memory-based adaptation, exercising autonomy rather than serving as passive executors of LLM guidance.  
After each cycle, agents report local actions, conditions, and outcomes, enabling the LLM to update its understanding. 
Final action choices remain agent-driven and are moderated by a trust mechanism: agents weigh their own policies against LLM suggestions, increasing trust when external input improves performance and decreasing it otherwise. 
This balance preserves independence while leveraging strategic guidance when beneficial.  
Agents also receive dual rewards: 
\begin{enumerate}[label=(\roman*)]
    \item \textit{Immediate feedback} from the environment in response to their local actions.
    \item \textit{Delayed feedback} from the \textit{Centralized Alignment Controller} for their contributions to broader strategic goals. 
\end{enumerate}
 This hierarchical reward mechanism ensures that the objectives of the individual agent remain aligned with the global network objectives while maintaining local autonomy in decision-making. This process is outlined in Algorithm~\ref{alg:train-agent}.

 \noindent {\textbf{Feedback Unit for Communication.}} 
To reduce latency in LLM–agent interaction, AURA employs batched communication: subgoal updates and feedback are aggregated at fixed intervals, lowering overhead while preserving responsiveness~\citep{zhuang2024yolo}. 
This limits continuous feedback during training, enabling agents to adapt locally while remaining aligned with global goals.  
To further improve adaptability, AURA incorporates Reflexion-inspired verbal feedback~\citep{Shinn2023ReflexionLA}. 
These lightweight updates provide corrective cues (e.g., reprioritizing subgoals during congestion) without costly retraining, guiding agents to switch strategies under high demand.  
A \textit{language-to-policy translator} bridges verbal instructions and MARL execution, using semantic parsing (inspired by GPT-like models~\citep{OpenAI2024}) to convert guidance, such as adjusting exploration or prioritizing goals, into actionable parameters that shape agents’ decision-making.


%% file: eval.tex
\section{Evaluation Scenario: Managing Network Overload Over Varying Traffic Conditions}

We evaluate AURA in a custom simulation with two base stations: a \textit{rural} station ($43 - 46$ dBm, max $50$ users) and an \textit{urban} station ($30 - 37$ dBm, max $30$ users). 
Users arrive and depart dynamically, each assigned a random signal strength ($ -120 $ to $-50$ dBm) and SNR ($0 - 30$ dB). 
 Episodes begin with randomized power levels, user counts, and channel conditions to ensure variability. Derived policies are used by frameworks as starting point during the testing phase. The evaluation includes three configurations: \textsc{MARL-Only}, where agents operate independently without external input; \textsc{Guided MARL}, where agents receive high-level suggestions from the LLM and selectively adopt them based on a learned trust mechanism; and \textsc{AURA}, which combines LLM guidance with delayed reward shaping to align local agent behavior with global objectives. Additional experiment details can be found in Appendix~\ref{setup}.



%% file: exp_result.tex
\section{Experimental Results}

\begin{figure}
\vspace{-1.2cm}
\hspace{-.5cm}
    \includegraphics[width=1.05\linewidth]{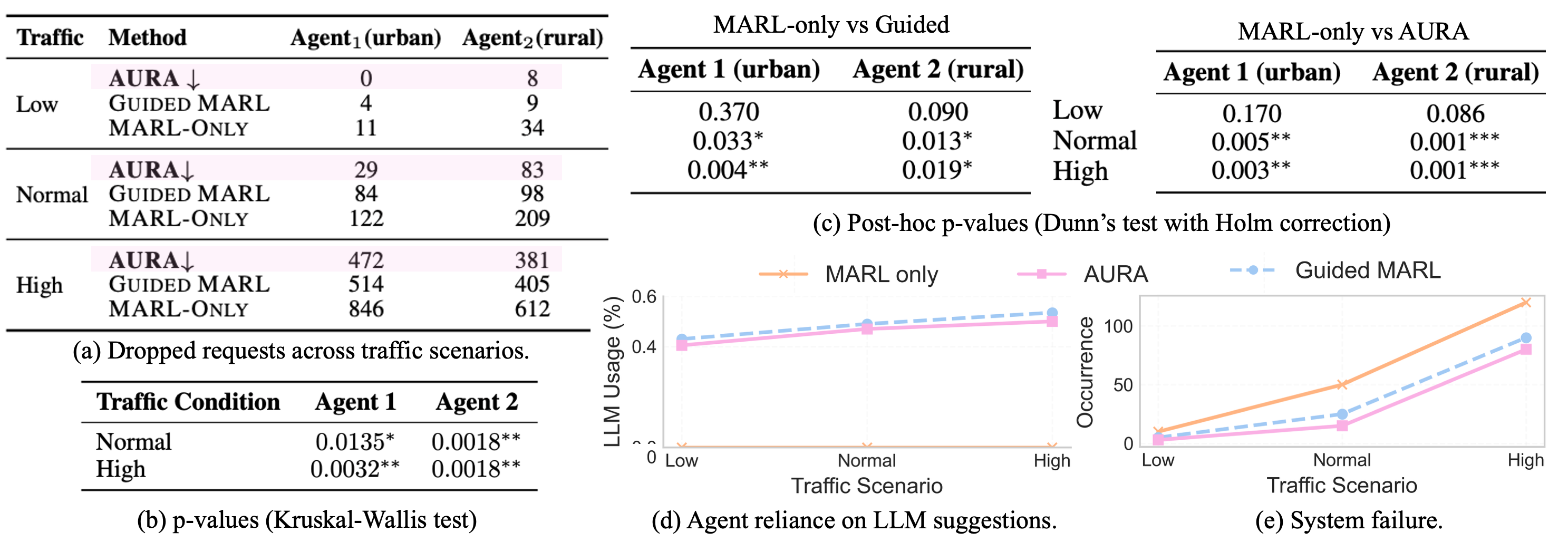}
    \caption{Comparison of \textsc{MARL-Only, Guided MARL}, and \textsc{AURA} across traffic conditions. 
    (a) Dropped requests, with largest gains under normal and high traffic.
(b–c) Statistical analysis showing strongest improvements for \textsc{AURA} and \textsc{Guided MARL} (* $p<.05$, ** $p<.01$, *** $p<.001$).
(d) LLM usage rates, indicating moderate reliance.
(e) System failure counts, reduced by both LLM-guided methods under higher traffic.}
    \label{main}
\end{figure}

We present preliminary results demonstrating the benefits of incorporating different levels of LLM guidance for adaptive network management under varying traffic conditions (Figure~\ref{main}). 
Both \textsc{Guided MARL} and \textsc{AURA} consistently reduce the number of dropped requests relative to \textsc{MARL-Only}, with the largest improvements under normal and high traffic (Table~\ref{main}a). 
As illustrated in Table~\ref{main}b significant difference\footnotemark[1] was observed for both agents under the normal and high traffic conditions when comparing number of dropped requests across all configurations (no significant differences were detected in low traffic). To further identify which methods drive these differences, we apply pairwise Dunn post-hoc tests against the \textsc{MARL-Only} baseline (Table~\ref{main}c -- stars denote threshold). These results confirm that \textsc{AURA} yields the strongest improvements ($p<0.01$), while \textsc{Guided MARL} also outperforms the baseline, albeit with weaker evidence. Taken together, these findings indicate that LLM guidance is most beneficial when the network is stressed, where MARL-only agents struggle to sustain reliable handoffs.

Figure~\ref{main}d further shows that agents rely on LLM suggestions only moderately (below 60\%), even in high-traffic scenarios, demonstrating that guidance does not lead to overdependence. \textsc{AURA} exhibits slightly lower reliance than \textsc{Guided MARL}, as delayed reward shaping provides additional signals that encourage agents to refine their policies more independently. This highlights that the framework can exploit LLM strategies while preserving local adaptability, thereby reducing latency from frequent queries, avoiding overdependence on a single centralized model, and mitigating errors from hallucinated or non-grounded guidance.
Finally, Figure~\ref{main}e reports system failures, defined as the number of testing steps in which the system failed to serving all requests, directly reflecting the frequency of system breakdowns. As expected, failure rates escalate with traffic intensity, but both \textsc{Guided MARL} and \textsc{AURA} substantially reduced these occurrences compared to \textsc{MARL-Only}, with \textsc{AURA} achieving the lowest failure counts under high traffic.

\footnotetext[1]{Since the normality assumption did not hold, we applied the non-parametric Kruskal–Wallis~\citep{kruskal1952} test to compare number of dropped calls across all configurations for each agent, followed by pairwise Dunn's test with Holm correction ~\citep{dinno2015nonparametric} to identify which configurations differed significantly.}

\section{Conclusion and Future Work}

This work introduced \textsc{AURA}, an LLM-guided MARL framework that balances strategic planning with localized adaptability in dynamic cellular networks. By combining LLM objectives with trust-gated action adoption and delayed reward shaping, the framework reduces failures and improves resilience under different traffic dynamics. Preliminary results show reduces failures and dropped requests under stress while limiting LLM reliance to <60\%. 
Future work includes extending the multimodal processor and LLM-based planner to provide richer context and strategic guidance. Key directions are scaling \textsc{AURA} to larger multi-agent deployments, improving robustness to imperfect LLM input through uncertainty estimation and trust calibration, enhancing explainability to support operator oversight, and benchmarking against emerging 6G baselines for realistic evaluation.

%% file: appendix.tex

\appendix


%% file: related_work.tex
\section{Related Work}
Artificial intelligence (AI) has been extensively used in cellular networks to address challenges in network optimization. 
In particular, supervised and unsupervised learning techniques, especially deep learning, have shown considerable potential to improve various aspects of network operations~\citep{zappone2019wireless, maseer2024meta}. 
Supervised learning has been widely utilized when it comes to predictive networking tasks, including traffic classification and bandwidth prediction. 
By training deep neural networks (DNNs) on historical data, these methods enable automated decision-making to optimize system performance~\citep{wang2020supervised, bentaleb2019bandwidth}. 
Nonetheless, the growing complexity and evolving demands of 6G networks expose fundamental limitations in these approaches. 
Common issues include scalability and limited adaptability to real-time fluctuations. Moreover, the diversity of networking tasks further limits model reuse, as each problem often requires a distinct architecture~\citep{ren2021comprehensive}. 
Recent advances, such as structured Transformers~\citep{vaswani2017attention}, have improved adaptability. Even so, they still rely on manual tuning and architectural modifications~\citep{wu2024mansy}, increasing the complexity of the deployment and engineering costs.

In response to the challenges of conventional learning approaches, researchers have increasingly turned to reinforcement learning (RL) to meet the dynamic requirements of modern networks~\citep{chen2018auto}. 
RL has been applied to various areas spanning congestion control, traffic optimization, and cloud cluster job scheduling (CJS)~\citep{abbasloo2020classic, casas2020intelligent, nourzad2024smart}. 
Multi-agent RL (MARL) extends these capabilities by enabling collaboration among multiple agents, offering improved scalability and robustness in distributed systems~\citep{yang2022dynamic, ning2024survey, bloembergen2015evolutionary}. 
However, while these techniques perform well in controlled settings, real-world deployments present significant challenges. Computational inefficiencies slow down processing, high-dimensional environments hinder convergence, and dynamic conditions with unpredictable user behaviors make adaptation difficult~\citep{iqbal2024intelligent, tuyls2012multiagent}.

Recent research has begun exploring the integration of MARL with Large Language Models (LLMs) to improve coordination and decision-making~\citep{sun2024llm, Tuyls2023marl}. 
Leveraging the advanced planning and reasoning capabilities of LLMs, researchers have demonstrated that these models can coordinate multiple agents within a network~\citep{zou2024genainet, yan2025hybrid, he2025generative}. On top of that, LLMs can uncover patterns in large datasets that traditional RL methods may overlook. 
These capabilities simplify collective decision-making and enable real-time adaptation to fluctuating user demands in dynamic network scenarios~\citep{huang2022languagezeroplanner}. 
Furthermore, by integrating multimodal data and anticipating future network states, LLMs can optimize resource allocation to enhance both operational efficiency and network robustness in 6G environments~\citep{luketina2019surveyrlllm, zou2023wireless, khoramnejad2025generative}. 
This hybrid approach effectively bridges critical gaps in adaptability and decision-making under dynamic conditions. 
Regardless, its practical application remains under-explored, particularly in the context of multi-agent collaboration and real-time resource optimization, where scalability and efficiency are key.

%% file: alg.tex
\section{Reward Algorithm}

\begin{algorithm*}
\caption{MARL Training with LLM Guidance}
\label{alg:train-agent}
\begin{spacing}{1.1} 
\begin{algorithmic}
\tcbset{
    colback=delayed!40,     
    colframe=delayed,    
    arc=0pt,                
    boxrule=0pt,            
    width=1.01\linewidth,       
    boxsep=0pt,             
    before skip=0pt,        
    after skip=0pt,         
    sharp corners,          
    left=5pt,               
    right=2pt,              
    enlarge left by=-5pt  
}
\begin{tcolorbox}
\Function{Delayed\_Reward}{$\mathcal{A}$}
    \For{each $agent_i \in \mathcal{A}$}
        \State $r_{delayed,i} \gets \textsc{Compute\_Delayed\_Reward}(\mathcal{H}_i)$
        \For{each $(s, a) \in \mathcal{H}_i$}  \Comment{Apply rewards uniformly to all past experiences in history}
            \State $Q(s, a) \gets Q(s, a) + \alpha \cdot r_{delayed,i}$  \Comment{$\alpha$  controls how much the delayed reward influences learning}
        \EndFor
        \State $\textsc{Clear\_History}(\mathcal{H}_i)$
    \EndFor
\EndFunction
\end{tcolorbox}
\For{$epoch$ in $\text{range}(N)$}
    \For{each $agent_i \in \mathcal{A}$}
        \State $a_{i, LLM} \gets \pi\textsc{.Action}(s_i, s_{-i})$ \Comment{ \textcolor{llm}{Action recommended to $agent_i$ by LLM $\pi$}}
        \State $a_{i, RL} \gets agent_i \textsc{.Action}(s_i)$ \Comment{\textcolor{marl}{Action taken by $agent_i$ based on local, real-time information}}
        \State $a_i \gets agent_i \textsc{.Combine\_Decision}(a_{i, LLM}, a_{i, RL}, \text{trustscore}_i)$
        \State $s'_i, r_{immediate, i}, \psi_i \gets \textsc{Take\_Action}(a_i)$ 
        \Comment{\(\psi_i\) captures environmental parameters like SNR, signal strength, etc.}
        \State $Q(s_i, a_i) \gets Q(s_i, a_i) + \alpha \left[ r_{immediate,i} + \gamma \max\limits_{a'} Q(s'_i, a') - Q(s_i, a_i) \right]$
        \Comment{\textcolor{marl}{Q-learning update using immediate rewards}}
        \State $agent_i \textsc{.Update\_Trust}(\pi, s_i, a_{i, LLM}, a_{i, RL}, r_{immediate,i}, a_i)$
        
    \EndFor
\EndFor
\end{algorithmic}
\end{spacing}
\end{algorithm*}

Algorithm~\ref{alg:train-agent} describes MARL training under LLM guidance, where agents combine local policies with LLM-suggested actions using a trust score. 
Immediate rewards update Q-values in real time, while accumulated histories enable periodic delayed rewards that reinforce alignment with high-level objectives.  

%% file: exp_setup.tex
\section{Experimental Setup} \label{setup}

 \subsection{{MARL formulation.}} We model the system as a Partially Observable Markov Decision Process (POMDP), which more accurately reflects the environment compared to a fully observable MDP~\citep{hausknecht2015deep}. The LLM has a global, though partial, understanding of the network’s state based on various data inputs (e.g., social media and network statistics). While the LLM has access to a broader set of information in comparison to individual agents, it still deals with partial observability since it doesn’t have complete information about the environment (like the exact state of every user or base station). Similarly, the MARL agents operate with partial observability, since each agent is limited to their local observations (e.g., signal strength, traffic load). The agents must make decisions with incomplete knowledge of the broader network. In this setup, LLM serves as a centralized decision-maker that breaks down high-level goals into subgoals for the agents to execute locally. In other words, the LLM centrally reasons and provides strategic guidance, making it more aligned with a centralized POMDP, where the central entity (LLM) orchestrates the strategy based on partial information.
We now turn to defining the state, action, and reward spaces for the MARL agents within the AURA system:

\vspace{1mm} - {\textsc{Observation Space}}: Each agent’s state space is defined by a set of local features that capture network conditions relevant to decision-making: (1) the transmission \textit{power level} of the agent’s cell tower, (2) the quality of \textit{network coverage} (categorized as good, fair, or poor), (3) the user \textit{connection status}, which indicates whether the tower has reached its maximum capacity, and (4) number of calls dropped from last time. Based on these observations, agents must select actions that optimize network performance while adapting to environmental changes. 

 - {\textsc{Action Space}}: The available action space consists of four primary decisions. Agents can (1) \textit{increase} transmission power when signal strength is weak and capacity allows. They can (2) \textit{decrease} power to reduce interference or conserve energy. If network conditions are stable, they may (3) \textit{maintain} the current state. Finally, they can initiate a (4) \textit{handoff} request when a user’s connection quality deteriorates, provided a neighboring tower offers a better alternative.

 - {\textsc{Reward}}: The reward function combines both \textit{immediate} and \textit{delayed} rewards. Immediate rewards are assigned at each time step, providing feedback based on local performance indicators such as connectivity quality and energy efficiency. Delayed rewards, on the other hand, are computed every few episodes by the CAC using a language model-based evaluation mechanism. The delayed reward function processes the agent’s historical state-action trajectory to derive a performance-based reward score, considering factors such as alignment with high-level objectives and overall network impact. This additional feedback mechanism helps agents optimize long-term behavior rather than overfitting to short-term gains.

\subsection{{LLM Query Design.}} 

 \vspace{1mm} - {\textsc{Centralized language model}}: In our implementation, a detailed prompt is constructed to guide the LLM in suggesting actions based on the network condition.. The prompt is composed by incorporating key metrics from both the neighboring and target cell towers, including power levels, coverage quality, and user connection status.  We use \textit{Claude-sonnet-4} as the underlying model, which generates responses based on these inputs. Then, a translator module extracts a clean numeric code from the LLM’s response, ensuring that only the intended output is used by the MARL framework. This approach allows the LLM to focus solely on processing the relevant state information and providing concise, actionable feedback, while the broader action execution is managed by the MARL framework.

 - {\textsc{Centralized alignment controller}}: As described earlier, the Centralized Alignment Controller (CAC) plays two roles in our framework. However, at this stage of our work, we focus on its function as a reward provider, evaluating MARL agents’ performance and assigning rewards accordingly.
 To implement this, we use an LLM-based evaluation mechanism that instructs the model, to act as an expert evaluator. The evaluation considers factors such as network efficiency, fairness, adaptability, and long-term performance. Based on these criteria, a reward is assigned within the range of $[-1, 1]$.  A score of +1 indicates "\textit{Excellent}" optimization (i.e., maximized efficiency, user experience, and adaptability), while -1 reflects "\textit{Poor}" optimization (i.e., significant issues like congestion, dropped connections, or poor resource use). Intermediate values represent varying levels of improvement or decline.

 Prompts and implementation details can be found in: https://anonymous.4open.science/r/AURA-F79F/.